\documentclass{appolb}
\usepackage{epsfig}

\begin{document}



\title{Static-static-light-light tetraquarks in lattice QCD\thanks{Presented at ``Excited QCD 2011'', 20--25 February 2011, Les Houches, France.}}

\author{Marc Wagner
\address{Humboldt-Universit\"at zu Berlin, Institut f\"ur Physik, Newtonstra{\ss}e 15, \\ D-12489 Berlin, Germany}
}

\maketitle

\vspace{-0.7cm}
\begin{center}
\includegraphics{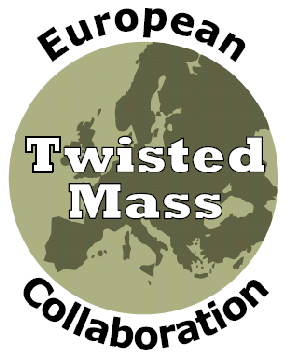}
\end{center}

\begin{abstract}
I report on a lattice computation of the energy of a system of two light quarks and two static antiquarks as a function of the separation of the static antiquarks. In terms of hadrons such a system corresponds to a pair of $B$ mesons and its energy to the hadronic potential. I present selected results for different isospin, spin and parity combinations of the individual $B$ mesons mainly focusing on those channels relevant to determine, whether two $B$ mesons may form a bound tetraquark state.
\end{abstract}




\section{Introduction}

Lattice computations of the potential of a pair of static-light mesons (in the following also referred to as $B$ mesons) are of interest, because they constitute first principles determinations of a hadronic force. Such potentials can e.g.\ be used as input for phenomenological calculations to determine, whether two $B$ mesons may form a bound tetraquark state.

In the literature interactions between static-light mesons have been studied in the quenched approximation \cite{Stewart:1998hk,Michael:1999nq,Cook:2002am,Doi:2006kx,Detmold:2007wk} and recently also with dynamical quarks \cite{Wagner:2010ad,Bali:2010xa}. Here I report on the status of an investigation with two flavors of dynamical Wilson twisted mass quarks. Forces are not only studied between the lightest static-light mesons (denoted by $S$), but for the first time also first excitations are taken into account (denoted by $P_-$).






\section{Trial states and quantum numbers}

Quantum numbers of single static-light mesons as well as of pairs of static-light mesons ($B B$ systems) have been discussed in detail in \cite{Wagner:2010ad}. In the following I give a brief summary.


\subsection{\label{SEC001}Static-light mesons}

I consider static-light mesons made from a static antiquark $\bar{Q}$ and a light quark $\psi \in \{ u \, , \, d \}$ without non-trivial gluonic excitations. They can be labeled by the $z$-component of isospin $I_z = \pm 1/2$, the $z$-component of the light quark spin $j_z = \pm 1/2$ and parity $\mathcal{P} = \pm$. The $\mathcal{P} = -$ static-light meson (denoted by $S$, corresponding to $B$/$B^\ast$ in \cite{PDG}) is the lightest static-light meson. It is lighter by around $400 \, \textrm{MeV}$ than its parity partner with $\mathcal{P} = +$ (denoted by $P_-$, corresponding to $B_0^\ast$ and $B_1^\ast$).

The corresponding static-light meson trial states are $\bar{Q} \gamma_5 \psi | \Omega \rangle$ and \\ $\bar{Q} \gamma_j \psi | \Omega \rangle$ for $S$ mesons and $\bar{Q} \psi | \Omega \rangle$ and $\bar{Q} \gamma_j \gamma_5 \psi | \Omega \rangle$ for $P_-$ mesons, respectively.

For a more detailed discussion of static-light mesons I refer to \cite{Jansen:2008si,:2010iv}.


\subsection{\label{SEC002}$B B$ systems}

The aim of this work is to compute the potential of a pair of $B$ mesons as a function of their separation $R$ (without loss of generality I choose the axis of separation to be the $z$ axis). To this end one has to compute the energies of eigenstates of the QCD Hamiltonian containing two static antiquarks $\bar{Q}(\mathbf{r}_1)$ and $\bar{Q}(\mathbf{r}_2)$, $\mathbf{r}_1 = (0,0,-R/2)$ and $\mathbf{r}_2 = (0,0,+R/2)$, which define the positions of the two $B$ mesons, and which will be surrounded by light quarks and gluons.

These $B B$ states are characterized by the following five quantum numbers: isospin $I \in \{ 0 \, , \, 1 \}$, the $z$-component of isospin $I_z \in \{ -1 \, , \, 0 \, , \, +1 \}$, the absolute value of the $z$-component of the light quark spin $|j_z| \in \{ 0 \, , \, 1 \}$, parity $\mathcal{P} = \pm$ and ``$x$-parity'' (reflection along the $x$-axis) $\mathcal{P}_x = \pm$.

I use $B B$ trial states
\begin{eqnarray}
\label{EQN001} (\mathcal{C} \Gamma)_{A B} \Big(\bar{Q}_C(\mathbf{r}_1) \psi_A^{(1)}(\mathbf{r}_1)\Big) \Big(\bar{Q}_C(\mathbf{r}_2) \psi_B^{(2)}(\mathbf{r}_2)\Big) | \Omega \rangle ,
\end{eqnarray}
where the lower indices $A$, $B$ and $C$ denote spinor indices, $\mathcal{C} = \gamma_0 \gamma_2$ is the charge conjugation matrix and $\Gamma$ is a suitably chosen combination of $\gamma$ matrices. Note that it is essential to couple the light degrees of freedom of both mesons in spinor space, because these degrees of freedom determine the quantum number $|j_z|$. Proceeding in a naive way by coupling light and static degrees of freedom in both $B$ mesons separately will not result in a well defined angular momentum $|j_z|$ and, therefore, will mix different $B B$ sectors. To obtain $I = 0$, the flavors of the light quarks have to be chosen according to $\psi^{(1)} \psi^{(2)} = u d - d u$, while for $I = 1$ three possibilities exist, $\psi^{(1)} \psi^{(2)} \in \{ u u \, , \, d d \, , \, ud + d u \}$. For a list of $B B$ trial states and their quantum numbers I refer to \cite{Wagner:2010ad}, Table~1.


\section{Lattice setup}

I use $24^3 \times 48$ gauge field configurations generated by the European Twisted Mass Collaboration (ETMC). The fermion action is $N_f = 2$ Wilson twisted mass \cite{Frezzotti:2000nk,Frezzotti:2003ni} at maximal twist, where static-light mass differences are automatically $\mathcal{O}(a)$ improved \cite{Jansen:2008si}. The gauge action is tree-level Symanzik improved \cite{Weisz:1982zw}. I use gauge coupling $\beta = 3.9$ and light quark mass $\mu_\mathrm{q} = 0.0040$ corresponding to a lattice spacing $a = 0.079(3) \, \textrm{fm}$ and a pion mass $m_\mathrm{PS} = 340(13) \, \textrm{MeV}$ \cite{Baron:2009wt}. For details regarding these gauge field configurations I refer to \cite{Boucaud:2007uk,Boucaud:2008xu}.



In twisted mass lattice QCD at finite lattice spacing SU(2) isospin is explicitely broken to U(1), i.e.\ $I_z$ is still a quantum number, but $I$ is not. Moreover, parity $\mathcal{P}$ has to be replaced by twisted mass parity $\mathcal{P}^{(\textrm{\scriptsize tm})}$, which is parity combined with light flavor exchange. The consequence is that there are only half as many $B B$ sectors in twisted mass lattice QCD as there are in QCD, i.e.\ QCD $B B$ sectors are pairwise combined. Nevertheless, it is possible to unambiguously interpret states obtained from twisted mass correlation functions in terms of QCD quantum numbers. The method has successfully been applied in the context of static-light mesons \cite{Blossier:2009vy} and is explained in detail for kaons and $D$ mesons in \cite{Baron:2010th}. For a more elaborate discussion of twisted mass symmetries in the context of $B B$ systems I refer to an upcoming publication \cite{MW2010}.


\section{Selected numerical results}

The potential of a pair of $B$ mesons is extracted from the exponential falloff of correlation functions of trial states (\ref{EQN001}). As explained in subsection~\ref{SEC002} the trial states differ by the spin coupling of the light quarks via the $4 \times 4$ matrix $\Gamma$ ($16$ possibilities) and by their light quark flavor ($4$ possibilities). Consequently, there are $16 \times 4 = 64$ different correlation functions. This number can also be understood from the point of view of individual $B$ mesons: since each of the two $B$ mesons has $8$ possibilities regarding quantum numbers ($I_z = \pm 1/2$, $j_z = \pm 1/2$, $\mathcal{P} = \pm$), there should be $8 \times 8 = 64$ $B B$ correlation functions.

\begin{figure}[htb]
\begin{center}
\includegraphics{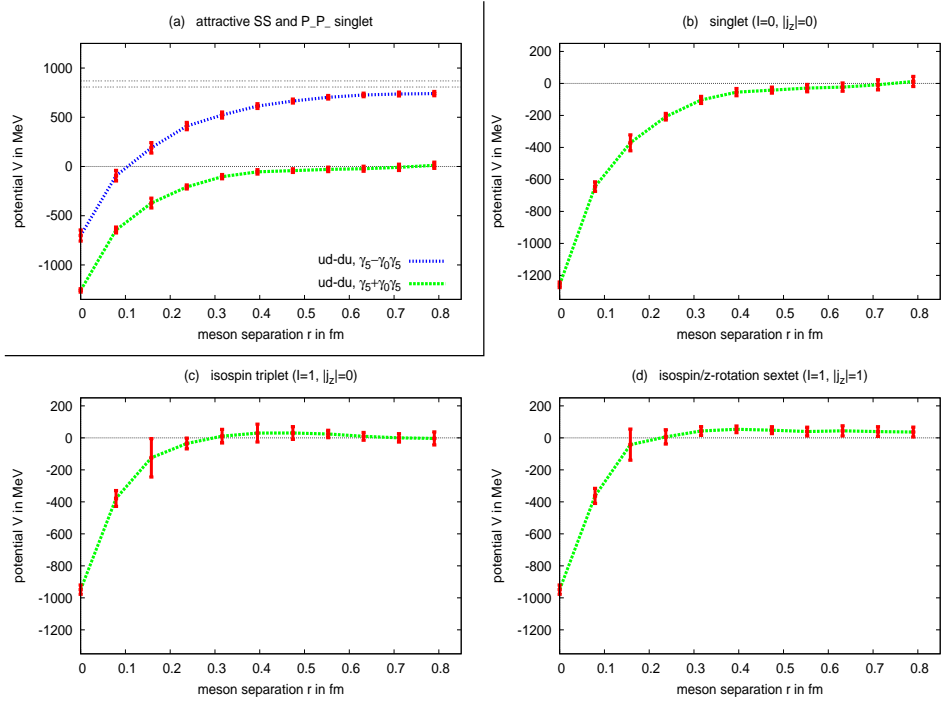}
\caption{\label{FIG001}$B B$ potentials as functions of the meson separation.
(a)~$S S$ potential and $P_- P_-$ potential both with quantum numbers $(I,I_z,|j_z|,\mathcal{P},\mathcal{P}_x) = (0,0,0,-,+)$.
(b),~(c),~(d)~Attractive $S S$ singlet, isospin triplet and isospin/spin sextet.
}
\end{center}
\end{figure}

As outlined in subsection~\ref{SEC002} the $B B$ trial states (\ref{EQN001}) can be classified according to $B B$ quantum numbers. However, to interpret the $B B$ potential obtained from the correlation function of a specific trial state (\ref{EQN001}), it is also useful to express this trial state in terms of individual $B$ mesons, e.g.\ \vspace{0.2cm} \\
$\psi^{(1)} \psi^{(2)} = ud - du \ , \ \Gamma = \gamma_5 \phantom{\gamma_0}$ $\quad \rightarrow \quad$ $(- S_\uparrow S_\downarrow + S_\downarrow S_\uparrow - P_\uparrow P_\downarrow + P_\downarrow P_\uparrow) | \Omega \rangle$ \\
$\psi^{(1)} \psi^{(2)} = ud - du \ , \ \Gamma = \gamma_0 \gamma_5$ $\quad \rightarrow \quad$ $(- S_\uparrow S_\downarrow + S_\downarrow S_\uparrow + P_\uparrow P_\downarrow - P_\downarrow P_\uparrow) | \Omega \rangle$, \vspace{0.2cm} \\
where $S$ and $P$ denote creation operators for $S$ and $P_-$ mesons, respectively, and $\uparrow$ and $\downarrow$ indicate the orientation of the light quark spin. After linearly combining the two trial states via $\Gamma = \gamma_5 + \gamma_0 \gamma_5$ to eliminate the $P_- P_-$ contributions, one can extract an $S S$ potential with quantum numbers $(I,I_z,|j_z|,\mathcal{P},\mathcal{P}_x) = (0,0,0,-,+)$. Similarly, one can estimate a $P_- P_-$ potential with the same quantum numbers by choosing $\Gamma = \gamma_5 - \gamma_0 \gamma_5$. Results are shown in Figure~\ref{FIG001}(a). Further examples are discussed in \cite{Wagner:2010ad}.

$B B$ potentials mainly differ in their asymptotic value at large meson separations, which is approximately $2 m(S)$, $m(S) + m(P_-)$ or $2 m(P_-)$ for $S S$, $S P_-$ or $P_- P_-$ combinations, respectively, and whether they are attractive or repulsive at small meson separations. Some of the potentials, even though they differ in their quantum numbers, are exactly degenerate due to isospin symmetry ($I = 1$ triplets) or rotational symmetry around the $z$-axis ($|j_z| = 1$ doublets). In summary the number of attractive and repulsive $B B$ potentials for $S S$, $S P_-$ and $P_- P_-$ combinations and their degeneracies are as follows: \vspace{0.2cm} \\
\begin{tabular}{llll}
$S S$ potentials,     & attractive: & $1 \oplus 3 \oplus 6$ & ($10$ states). \\
                      & repulsive:  & $1 \oplus 3 \oplus 2$ & ($\phantom{0}6$ states). \vspace{0.1cm} \\
$S P_-$ potentials,   & attractive: & $1 \oplus 1 \oplus 3 \oplus 3 \oplus 2 \oplus 6$ & ($16$ states). \\
                      & repulsive:  & $1 \oplus 1 \oplus 3 \oplus 3 \oplus 2 \oplus 6$ & ($16$ states). \vspace{0.1cm} \\
$P_- P_-$ potentials: & \multicolumn{2}{l}{identical to $S S$ potentials} & ($16$ states). \vspace{0.2cm}
\end{tabular} \\
Consequently, from the $64$ trial states (\ref{EQN001}) one can extract $24$ different potentials.

Of particular interest, when asking, whether two $B$ mesons may form a bound tetraquark state, are attractive $S S$ potentials. As stated above there are three different possibilities, a singlet, an isospin triplet and an isospin/spin sextet. The three potentials are shown in Figure~\ref{FIG001}(b) to \ref{FIG001}(d). While the triplet and the sextet are rather similar, the singlet is significantly different: it is deeper and wider and, therefore, probably the best candidate to start investigations, whether there are bound $B B$ tetraquark states. Such phenomenological investigations will be part of a subsequent publication.


\section{Conclusions}

I have presented selected results of a first principles lattice computation of $B B$ potentials. Various channels characterized by the quantum numbers $(I , I_z , |j_z| , \mathcal{P} , \mathcal{P}_x)$ have been investigated. Compared to existing publications the computations have been performed with rather light dynamical quark mass ($m_\mathrm{PS} \approx 340 \, \textrm{MeV}$). The results have been interpreted in terms of individual $S$ and $P_-$ mesons. Future plans include studying the light quark mass dependence, the continuum limit as well as finite volume effects. Moreover, also $B B_s$ and $B_s B_s$ potentials could be computed. To treat the $s$ quark as a fully dynamical quark, such computations should be performed on $N_f = 2+1+1$ flavor gauge field configurations currently generated by ETMC \cite{Baron:2010bv}.
Finally, one should use the obtained $B B$ potentials as input for phenomenological considerations to answer e.g.\ the question,  whether two $B$ mesons are able to form a bound tetraquark state.


\section*{Acknowledgments}

I thank the organizers of ``Excited QCD 2011'' for the invitation to give this talk. I acknowledge useful discussions with Pedro Bicudo, William Detmold, Rudolf Faustov, Roberto Frezzotti, Vladimir Galkin, Chris Michael and Attila Nagy. This work has been supported in part by the DFG Sonderforschungsbereich TR9 Computergest\"utzte The\-o\-re\-tische Teilchenphysik.



\end{document}